\documentclass{emulateapj}

\slugcomment{Submitted: \today}
\shorttitle{Integral-Field Spectroscopy of G11.2$-$0.3}
\shortauthors{Lee et al.}

\begin{document}

\def\kms{km~s$^{-1}$}
\def\um{$\mu$m}
\def\gele{G11.2$-$0.3}
\def\feii{[\ion{Fe}{2}]}
\def\hei{\ion{He}{1}}
\def\simlt{\lower.5ex\hbox{$\; \buildrel < \over \sim \;$}}
\def\simgt{\lower.5ex\hbox{$\; \buildrel > \over \sim \;$}}
\def\ug{$\mu$G}
\def\msol{{$M_\odot$}}
\def\lsol{{$L_\odot$}}
\def\rsol{{$R_\odot$}}
\def\feratio{\feii~1.534 to 1.644 $\mu$m}

\title{
Wide Integral-Field Infrared Spectroscopy of the Bright \feii\ Shell in the Young Supernova Remnant G11.2$-$0.3
}

\author{Ho-Gyu Lee\altaffilmark{1,2},
Dae-Sik Moon\altaffilmark{2,3}, 
Bon-Chul Koo\altaffilmark{4},
Mubdi Rahman\altaffilmark{2,5},
Stephen S. Eikenberry\altaffilmark{6},
Nicolas Gruel\altaffilmark{6},
Takashi Onaka\altaffilmark{1},
Hyun-Jeong Kim\altaffilmark{4},
Won-Seok Chun\altaffilmark{4},
John Raymond\altaffilmark{7}, 
S. Nicholas Raines\altaffilmark{6}, and
Rafael Guzman\altaffilmark{6}
}

\altaffiltext{1}{Department of Astronomy, Graduate School of Science,
The University of Tokyo, Tokyo 113-0033, Japan;
hglee, onaka@astron.s.u-tokyo.ac.jp}
\altaffiltext{2}{Department of Astronomy and Astrophysics, 
University of Toronto, Toronto, ON M5S 3H4, Canada; 
moon@astro.utoronto.ca}
\altaffiltext{3}{Visiting Brain Pool Scholar, Korea Astronomy and Space Science Institute, 
776 Daedeok-daero, Yuseong-gu, Daejeon 305-348, Republic of Korea}
\altaffiltext{4}{Department of Physics and Astronomy,
Seoul National University, Seoul 151-742, Republic of Korea;
koo, hjkim@astro.snu.ac.kr}
\altaffiltext{5}{Department of Physics and Astronomy, Johns Hopkins University, 
Baltimore, MD 21218, USA; mubdi@pha.jhu.edu}
\altaffiltext{6}{Department of Astronomy, 
University of Florida, FL 32611-2055, USA; eiken, raines@astro.ufl.edu}
\altaffiltext{7}{Harvard-Smithsonian Center for Astrophysics, 
60 Garden Street, Cambridge, MA 02138, USA; jraymond@cfa.harvard.edu}

\begin{abstract}
We present the results of wide integral-field near-infrared (1.0--1.8~\um) 
spectroscopic observations of the southeastern shell of the young core-collapse 
supernova remnant (SNR) G11.2$-$0.3.
We first construct \feii~1.644~\um\ line images of three bright clumps from the
obtained spectral image cubes and compare them with those of other transitions
such as \feii~1.257, \feii~1.534 and \hei~1.083~\um\ line images. 
This allows us to estimate the electron density ($\sim$ 4,700--9,400 cm$^{-3}$) 
and extinction ($A_{\rm V}$ $\sim$ 16--20 mag) of the shell, 
including detailed two-dimensional distribution of the properties in the brightest clump,
as well as the discovery of a faint high-velocity ($\sim$ --440 \kms) component in the clump.
Our SNR shock model calculations estimate the preshock number density of 
$\sim$ 250--500 cm$^{-3}$ and shock speed of $\sim$ 80--250 \kms\ in the \feii-emitting region of the SNR.
The comparison between the observed and modelled radial profiles of 
the line intensities and their ratios reveals that the shell is 
composed of multiple thin filaments which have been likely formed 
in episodic mass loss processes of a progenitor star.
The discovery of the faint high-velocity component supports the interpretation that
the southeastern shell of \gele\ is mainly composed of circumstellar material 
with contamination by supernova ejecta and also that 
its ejected material was expelled 
primarily in the southeast-northwest direction.
\end{abstract}

\keywords{ISM: individual objects (\objectname{G11.2$-$0.3}) --- 
ISM: supernova remnants ---
infrared: ISM ---
shock waves
}

\section{Introduction}

Near-infrared \feii\ observation of young core-collapse SNRs 
in the Galactic plane 
is an invaluable tool for understanding many interesting phenomena associated with 
core-collapse supernova explosions.
Iron is the end product of the stellar nucleosynthetic processes with 
relatively small (7.90 eV) ionization potential, 
which renders near-infrared \feii\ observations particularly useful for 
studying the distribution and kinematics of ejecta from deep layers 
of a progenitor star and those of circumstellar material from its later evolutionary phases
prior to the explosion, along with interstellar shocks, 
while suffering less extinction effect than the visual or X-rays.

An enlightening illustration is the heavily extinquished ($A_{\rm V}$ $\simgt$ 13 mag) 
young core-collapse SNR \gele, the likely remains of the historical supernova explosion in SN 386 
at $\sim$ 5 kpc distance \citep{ste02, bec85, gre88},
of which recent \feii\ observations have revealed several pertinent features.
According to the results of \citet{koo07} and \citet{moo09},
\gele\ is filled with bright near-infrared \feii\ emission 
originating from both the shocked ejecta and circumstellar material
scattered around the entire remnant.
In its inner part, knotty \feii\ emission with significant (\simgt\ 1,000 \kms)
Doppler shifts is prominent without any other bright elements,
indicating that the majority of the iron knots therein 
originate 
deep inside its progenitor.
Near its northwestern boundary lie knotty filaments of \feii\ emission,
reminiscent of bubbly iron ejecta.
The most conspicuous near-infrared \feii\ feature in \gele, however,
is its southwestern shell which is in fact the brightest near-infrared \feii\
emission ever detected among all Galactic SNRs.
The shell is thick and prominent in radio, X-ray and mid-infrared emission too \citep{gre88, kas01, pin11}.
Although all these are suggestive that the southwestern shell is a rare example
of shocked supernova ejecta interacting with shocked circumstellar material,
the origin and nature of this unique feature are uncertain.

One efficient way to investigate the \feii\ emission of the southwestern shell
of \gele\ is to obtain broadband two-dimensional integral-field spectra of the entire emission
which can provide both spectral and spatial information simultaneously.
In this paper, we report the results of such observations using our own instrument
the Florida Image Slicer for Infrared Cosmology and Astrophysics (FISICA).
We describe our instrument setup and observations in \S~2, 
followed by results of the observations (\S~3) and discussions
focused on shock modelling of multiple near-infrared \feii\ line emission (\S~4). 
We summarize our results in \S~5.

\section{Instrument Setup and Observations}

The instrument setup used for the observations we report in this paper
is a combination of Florida Multi-object Imaging Near-infrared Grism Observational Spectrometer (FLAMINGOS) 
and FISICA.
The former is a facility instrument at the Kitt Peak National Observatory Mayall 4~m telescope.
One unique feature of this instrument is a small multi-object spectrograph (MOS) dewar,
separated from its main dewar, dedicated for cycling multi-object slit masks \citep{els98}.
We replaced the MOS dewar with FISICA which is an image slicer-based integral-field unit
composed of three monolithic mirror arrays \citep{eik06}. 
FISICA was developed by some of the authors of this paper 
as an accessory module for FLAMINGOS, and 
the combination of FLAMINGOS and FISICA provides an unprecedentedly large integral-field of view of 
$\sim$ 16\arcsec\ $\times$ 33\arcsec\ in the near-infrared regime.
This setup is uniquely ideal for studying kinematics and chemistry of extended objects in the Galactic plane,
such as the \feii\ emission of the southwestern shell of \gele\ as we report here.

The observations took place on 2006 June 17 and 18 using the $JH$-band grism of FLAMINGOS,
which covers the $\sim$ 1.0--1.8 $\mu$m range populated with many \feii\ emission lines
with $R$ $\simeq$ 1,300 spectral resolving power.
We observed three bright \feii\ clumps (which we call Clump~1, 2, and 3) 
within the southeastern shell of \gele\
originally detected in the \feii\ 1.644 $\mu$m narrow-band imaging observations 
\citep[][see Figure~\ref{fig_limg}]{koo07}.
We followed each exposure of 300 seconds toward the clumps with the same exposure toward nearby sky 
without any apparent \feii\ emission in order to subtract out sky background emission (including OH lines).
The total on-source integration times were 1,500, 2,100, and 900 seconds for Clump 1, 2, and 3, respectively,
and the same integration times were used for sky observations.
After flat fielding, we obtained wavelength solutions using HeNeAr arc spectra,
and used the standard star of SAO 187086 (G6V) for telluric calibration \citep{coe05}.
The astrometric solutions of the obtained data cubes are consistent with the known positions of stars 
located inside the integral fields of the clumps.

\section{Results}

Figure~\ref{fig_limg} presents pure \feii\ (1.644 \um) line images of Clumps 1, 2, and 3,
together with those of \hei\ (1.083 \um) and \feii\ (1.257 and 1.534 \um) of Clump 1.
To construct the line images,
we measure the intensities of the line transitions at every spaxel by 
fitting their profiles with a gaussian function.
The \feii\ (1.644 \um) line images of the three clumps
show extended ($>$ 25\arcsec) elongated structures along the southwestern shell of \gele.
For the brightest Clump~1, in addition to the 1.644 \um\ transition, 
we also detect significant emission of \feii\ from the transitions at 1.257 and 1.534 \um. 
The images of the latter transitions closely resemble that of the former,
with correlation coefficients of 0.94 (1.256 vs. 1.644 \um\ images) and 0.95 (1.534 vs. 1.644 \um\ images).
We also detect \hei\ (1.083 \um) line emission from Clump 1.
Although faint, its peak position is coincident with those of the \feii\ lines,
and it appears to extend in the same direction as the extension of the \feii\ images.

Figures~\ref{fig_sp}a and 2b show $JH$-band spectra of Clumps 1, 2, and 3 
integrated over the entire region of each clump.
As in Figure~\ref{fig_sp}, they are exclusively dominated by \feii\ transitions,
and in total we identify 12 \feii\ transitions.
The only line which is detected other than the \feii\ lines is \hei\ (1.083 \um) from Clump 1.
Table~\ref{tab_obs} contains the details of the detected line transitions
including their relative intensities.
The detected lines do not show any apparent Doppler shifts,
which is consistent with the results of \citet{moo09} around this area of the SNR,
indicating that their motions are mostly tangential.
However, in the profile of the \feii\ (1.644 \um) transition from Clump 1 
(Figure~\ref{fig_sp}c),
we identify an additional velocity component which we call 
the southeastern blueshifted high-velocity feature (or SE-BHVF).
The extension of SE-BHVF is $\sim$ 10\arcsec\ and its peak position is 
located at $\sim$ 15\arcsec\ from the main peak position of Clump 1 
in the northeastern direction (Figure~\ref{fig_limg}).
Using a two-velocity component gaussian fit of the profile,
we estimate the radial velocity of SE-BHVF to be --440 $\pm$ 100 \kms,
highly blueshifted from the systematic velocity ($+$45 \kms) of 
G11.2$-$0.3 \citep{gre88}.
The integrated intensity of SE-BHVF is $\simlt$ 2~\% of the main component.

\begin{figure}
\center
\includegraphics[scale=.5]{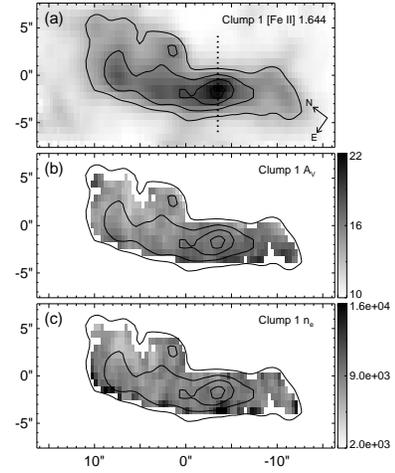}
\caption{
Distribution of the \feii~1.644 $\mu$m line intensity (a), extinction (b), and electron density (c) 
estimated in Clump~1.
Overlapped contours indicate \feii\ 1.644~$\mu$m line intensities at
30, 50, 70, and 90~\% levels of the peak value.
Vertical dotted line in (a) shows the positions along which
we obtain one-dimensional radial profiles of the \feii~1.644 $\mu$m 
line intensity and the \feratio\ line ratio
(Figure~\ref{fig_mo}).
}
\label{fig_par}
\end{figure}

Because the two \feii\ transitions at 1.257 and 1.644~\um\ detected in Clump~1 share the same upper level,
their intrinsic intensity ratio is determined by their atomic parameters.
We compare the observed ratio with 
the intrinsic ratio of 1.36 expected from the atomic parameters of \feii\ 
\citep{deb11} and compute the extinction toward the source
using the extinction curve of grains in the Milky Way with $R_{\rm V}$ = 3.1 \citep{dra03}.
Figure~\ref{fig_par}b shows a visual extinction ($A_{\rm V}$) map of Clump~1
which presents two-dimensional distribution of the extinction in the clump.
Its mean extinction is $A_{\rm V}$ = 16 $\pm$ 1 mag, but varies somewhat significantly
within the clump between 12 and 20 mag and increases in the southwestern directions
where Clumps 2 and 3 are located.
For the extinctions toward Clumps 2 and 3, we only calculate their average
extinctions of $A_{\rm V}$ = 18 $\pm$ 1 (Clump 2) and 20 $\pm$ 1 (Clump 3) mag
due to relatively weak intensities of their \feii\ 1.257 \um\ transitions.
The internal extinction distribution within Clump 1
and the increased extinctions toward Clumps 2 and 3 
show a tendency that the overall extinction toward G11.2$-$0.3 in this area
increases along the southern direction,
indicating potential existence of a foreground or circumstellar gas cloud there. 
The extinction toward Clump 1 is consistent
with a SN at 5 kpc being visible to the naked eye in AD 386.
The average extinction ($A_{\rm V}$ = 16 mag) of Clump 1 is higher than
the extinction ($A_{\rm V}$ = 13 mag) 
previously estimated at the intensity peak position 
\citep{koo07},
although their observed transition ratios of \feii\ 1.257 to 1.644 \um\ 
are similar within errors 
(0.303 $\pm$ 0.019 for the former and 0.314 $\pm$ 0.010 for the latter),
simply due to the different atomic parameters
and resulting intrinsic intensities 
used to estimate the extinctions \citep{deb11, qui96}.

We also use the distribution of the transition ratio of \feii~1.533 to 1.644~\um\ 
to obtain two-dimensional distribution of the electron density in Clump 1 (Figure~\ref{fig_par}c).
We solve the rate equations of 16 levels of Fe$^+$
using the transition probabilities and collision strengths 
in the literature \citep{deb11, ram07} at the temperature of 5,000~K. 
As in Figure~\ref{fig_par}c,
the distribution of the electron density in Clump 1 is totally different 
from that of the \feii\ 1.644 line intensities
with their peaks located at different positions.
The average electron density of Clump 1 is 9,400 $\pm$ 2,100 cm$^{-3}$,
while those of the other two clumps are
8,900 $\pm$ 2,800 and 4,700 $\pm$ 1,000 cm$^{-3}$
for Clumps 2 and 3, respectively.

\section{Discussion}

\begin{figure}
\center
\includegraphics[scale=.5]{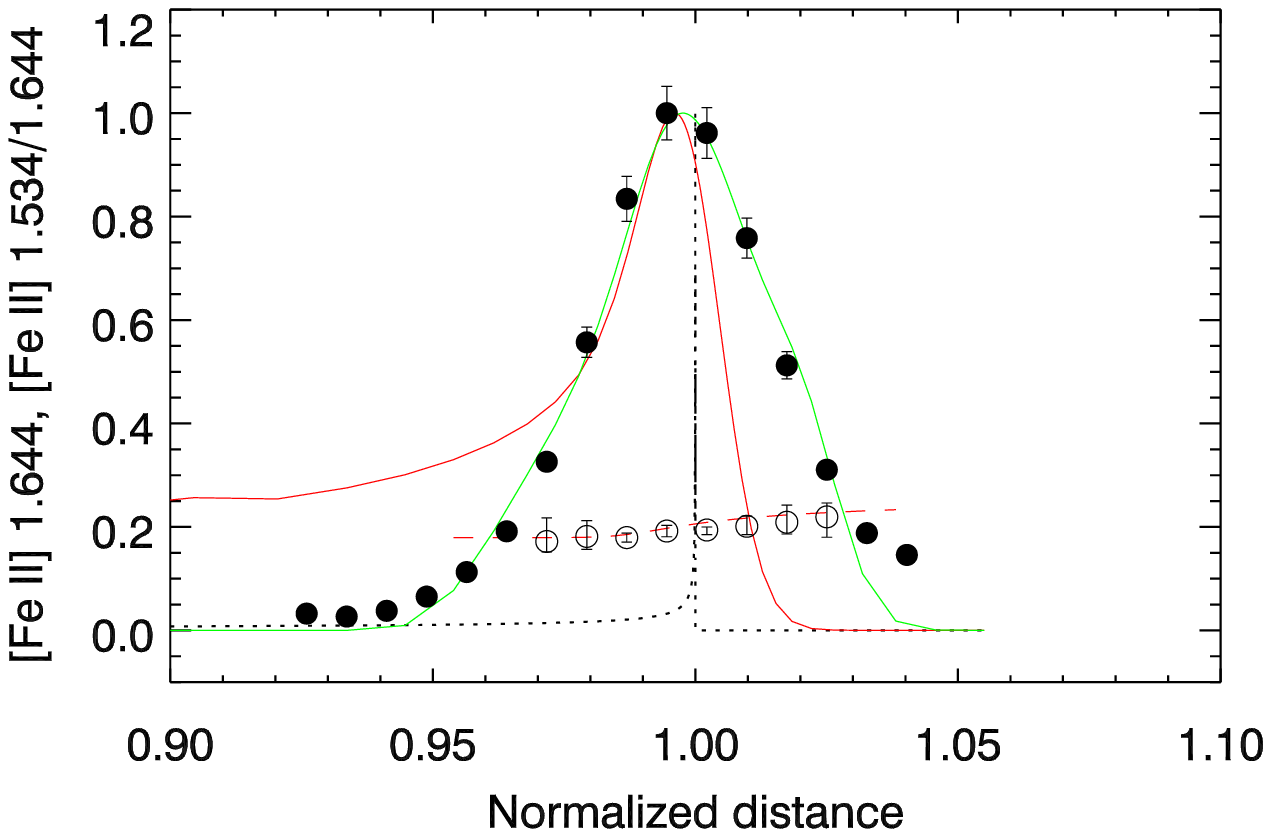}
\caption{
One-dimensional radial profiles of the \feii~1.644 \um\ intensity (filled circles)
and the \feratio\ line ratio (empty circles) along the cut crossing
the peak position of Clump 1 (see Figure 3).
The abscissa is the distance from the central pulsar of \gele\
normalized by the distance to the shock front.
The \feii~1.644 \um\ line intensity is normalized by its maximum intensity.
Also shown are model predicted profiles (see text for the details of the model):
black-dotted line for \feii~1.644 \um\ profile before beam dilution;
red-solid line for the same \feii~1.644 \um\ profile but after beam dilution;
red-dashed line for beam-diluted \feratio\ line ratio profile.
The green-solid line represents a beam-diluted \feii~1.644 \um\ profile
from the model calculation of multiple filaments of finite sizes.
}
\label{fig_mo}
\end{figure}

The near-infrared $JH$-band spectra of the observed clumps are dominated 
by \feii\ lines -- only weak emission of \hei\ is detected other than \feii\ lines
in Figure~\ref{fig_sp}.
One notable feature is the lack of correlation between the distribution
of the \feii\ line intensities and that of the electron number density
(Figure~\ref{fig_par}), which indicates that the \feratio\ ratio
distribution is largely independent of the line intensity distribution.
In order to investigate this more thoroughly, 
we compare the radial distribution of the two quantities,
i.e., \feii\ 1.644~$\mu$m line intensity and the \feratio\ line ratio, 
along the peak position of Clump~1 (Figure~\ref{fig_par})
by applying boxcar average of 5\arcsec.
Figure~\ref{fig_mo} shows that the radial distributions
of the two quantities are clearly dissimilar:
the \feii~1.644~$\mu$m line intensity peaks at the middle of the clump
and decays both inward and outward with $\sim$ 4\arcsec\ FWHM, 
while the \feratio\ ratio distribution is almost flat 
with a tendency of slight increase toward the outer edge of the clump.

We model the radial profiles of the \feii\ 1.534 and 1.644 \um\ line 
intensities and their ratio for a SNR shock penetrating into dense ambient medium
using a shock code (Figure~\ref{fig_mo}). 
The code that we use was
developed by \citet{raymond1979} and improved by \citet{cox1985}.  
We have updated the atomic parameters of \feii\ forbidden lines,
using recent values from \citet{deb11} and \citet{ram07} 
for the radiative transition rates and the collisional strengths, respectively. 
We calculate the level populations
of the lowest 16 levels in four terms ($a^6D$, $a^4F$, $a^4D$, and $a^4P$) 
of Fe$^+$ and the intensities of the lines
resulting from the transitions among them.
The line emissivities of a fluid element are calculated
following its trajectory from the shock front until it cools down to 1,000 K,
and are used to compute the line intensities and their ratios.
We assume that incoming H atoms are fully ionized by radiative
precursor. The code, however, does not calculate the emission from
radiative precusor itself. This is acceptable for \feii\ lines because
Fe atoms in the pre-shock gas may be mostly locked up to grains, which
are liberated when the grains are destroyed in the post-shock flow.
The \feii\ 1.534 and 1.644 $\mu$m line brightnesses and their ratio
are functions of shock speed ($v_s$),
preshock density ($n_0$), and preshock magnetic field strength ($B_0$)
perpendicular to the shock front in general.
We will present some general characteristics
of the \feii\ emission from radiative shocks using the shock code
in the forthcoming paper (Koo et al. in preparation).
Overall the preshock density is well constrained by the \feratio\ ratio.

The average \feratio\ ratio of Clump 1 is $0.192 \pm 0.012$ (Table~\ref{tab_obs}),
indicating the average postshock electron density
of $\sim 1 \times 10^4$~cm$^{-3}$ (see \S~3).
This value should be much higher than the uncompressed preshock density 
for a radiative shock.
For shock speeds of 80--250 \kms\ and $B_0=10$~\ug, the observed ratio implies
$n_0=250$--500 cm$^{-3}$. For $B_0=1$~\ug, it becomes $n_0=150$--300 cm$^{-3}$ for the same
shock speed range.
We adopt 10 \ug, which is the median value of the maximum magnetic field
strength for diffuse clouds \citep{crutcher2010}.
We take the shock speed to explain the observed line emission:
if the velocity is considerably lower,
e.g., \simlt\ 50 \kms, we may expect to see H$_2$ emission from molecular
shocks instead of bright \feii\ emission from atomic radiative shocks. 
If the velocity is siginificantly larger, e.g., $>$ 250 \kms, 
the expected brightness normal to the shock front becomes brighter than the observed one,
i.e., 1--2 $\times 10^{-2}$ ergs cm$^{-2}$ s$^{-1}$ sr$^{-1}$,
which is the intensity of the bright 
structure of Clump~1 obtained by \cite{koo07}
after being smoothed to the beamsize ($2''$)
of the current observation and dereddended using $A_V=16$~mag.
Note that the observed brightness is the integrated intensity 
along the line of sight and also beam-diluted.
The brightness along the line of sight is much larger than the brightness
normal to the shock front in general,
so that it is not straightforward to give more stringent constraints 
without more complicated and detailed shock model calculations.

\begin{figure*}
\center
\includegraphics[scale=.8]{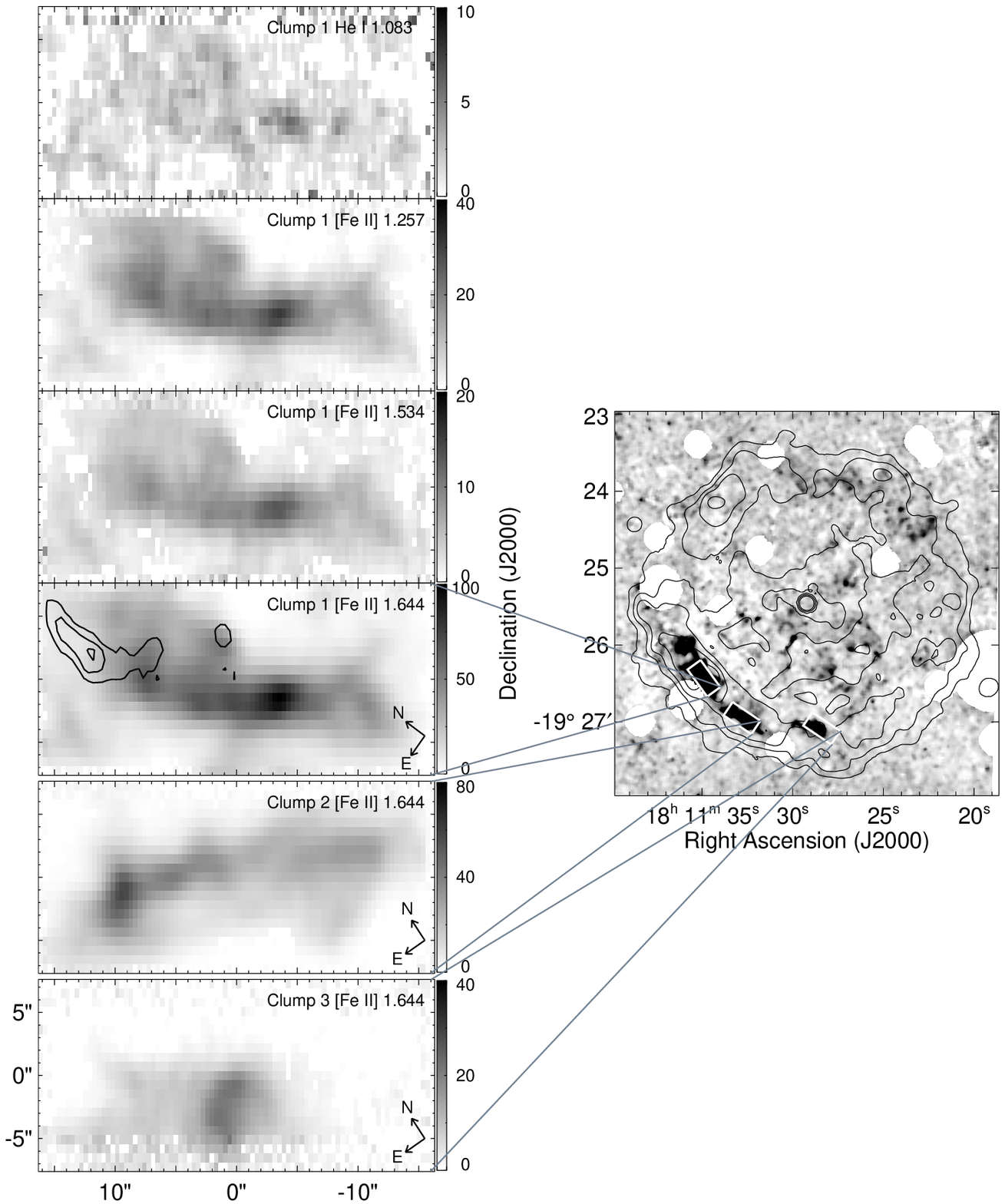}
\caption{
{\it Left}: 
Near-infrared line images of three bright \feii\ clumps within 
the southeastern shell of G11.2$-$0.3.
From the top, the gray-scale images are \hei~1.083, \feii~1.257, \feii~1.534, 
and \feii~1.644~\um\ line images of Clump~1, and 
\feii~1.644~\um\ line images of Clump~2 and Clump~3.
Gray-scale units are normalized to have peak intensities of 100
for the \feii~1.644~\um\ line image of Clump 1.
Contours on the \feii~1.644~\um\ line image of Clump~1
indicate the distribution of the SE-BHVF.
{\it Right}: 
\feii~1.644~\um\ narrow-band image of G11.2$-$0.3
overlaid on the Chandra X-ray contours \citep{koo07}.
Boxes represent the locations of the three clumps observed in this study.
}
\label{fig_limg}
\end{figure*}

\begin{figure}
\center
\includegraphics[scale=.4]{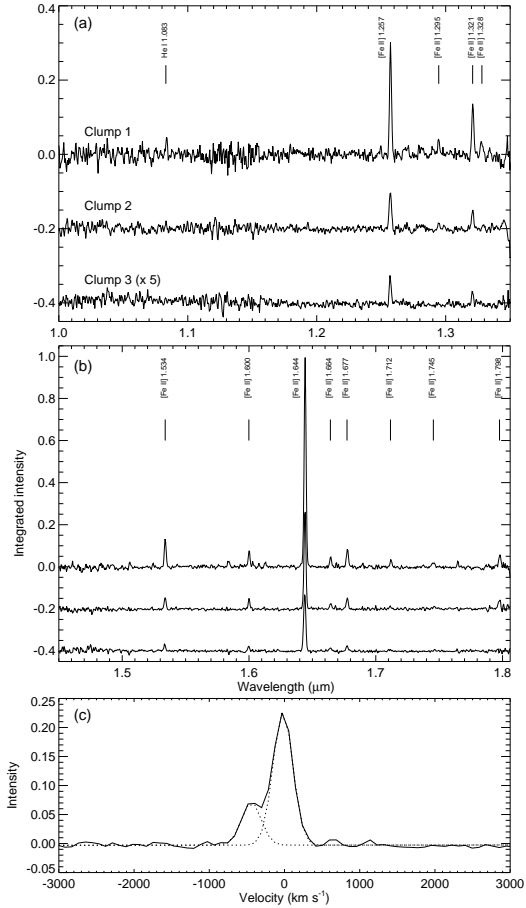}
\caption{
Integrated $J$- (a) and $H$-band (b) spectra of Clump 1, Clump 2, and Clump 3
over the entire region of each clump.
The intensities are normalized by the peak \feii~1.644~\um\ intensity of Clump 1.
Constant offsets of 0.2 and 0.4 are subtracted from 
the spectra of Clump 2 and Clump 3, respectively, 
in order to show their spectra effectively.
The spectra of Clump 3, which are faint, are multiplied by 5 for better display.
In the bottom panel (c), the solid line is the observed \feii~1.644~\um\ spectrum 
of SE-BHVF (see Figure~\ref{fig_limg} for the position),
whereas the dotted lines are two fitted gaussian components 
of --20 and --440 \kms.
}
\label{fig_sp}
\end{figure}

We perform a shock model calculation of the observed radial profiles 
assuming solar abundance \citep{asp09}.
In Figure~\ref{fig_mo}, the red-solid and red-dashed lines represent the normalized 
\feii\ 1.644 \um\ intensity and the \feratio\ line ratio, respectively,
for a spherically symmetric shock of 130~\kms\ propagating 
into an ambient medium of $n_0=300$~cm$^{-3}$. 
The peak intensity of \feii\ 1.644 \um\ line profile is 
8 $\times$ $10^{-3}$ ergs cm$^{-2}$ s$^{-1}$ sr$^{-1}$.
Note that these are profiles integrated along the line of sight and beam-diluted. 
For comparison, we also show the normalized \feii\ 1.644 \um\ intensity profile 
(black-dotted line) which is free of the beam dilution effect,
where we can identify a sharp intensity contrast caused by a thin shock front.
As in Figure~\ref{fig_mo}, our shock model fits the \feratio\ ratio very well, 
whereas it does the \feii\ 1.644 \um\ intensity profile somewhat poorly. 
The observed intensity (filled circles) is much fainter than the modelled one (red-solid line)
at $R\le 0.95$ and the width of the observed shell is much greater than what is modelled.
These discrepancies imply that the structure is not spherically symmetric,
with local enhancement of high-density medium in the region, and also that 
there may be multiple filaments inside the beam. 
We can fit the observed intensity profile and the peak brightness readily
using multiple filaments of finite sizes. 
For example, the green-solid line in Figure~\ref{fig_mo}  
shows the modelled intensity profile when there are three filaments of
1.7 pc size separated by 0.03--0.04 pc from each other,
and it is in a good agreement with the observed values.
The current model is probably not a unique model for the radial profile, 
but multiple filaments seem to be unavoidable. 
A possible scenario for the origin of such multiple filaments is 
an episodic mass loss from the progenitor of \gele\ 
to the southeast-northwest direction.
The similar distribution of \hei\ in Clump 1 (Figure~\ref{fig_limg})
also agrees with the circumstellar origin of the main \feii\ emission of the clump,
excluding SE-BHVF (see below).
The dereddended line ratio of \hei\ 1.083 to \feii\ 1.644 $\mu$m
is $\sim$ 0.50, 
which is comparable to $\sim$ 0.59 
previously observed in RCW~103 \citep{oli90}.

The three \feii\ clumps that we observe here
are surrounded by hot X-ray gas (Figure~\ref{fig_limg}). 
The density of the X-ray emitting gas is 13 cm$^{-3}$,
if we simply divide 
its ionization time scale $n_e t$ = 6.7 $\times$ 10$^{11}$ s cm$^{-3}$ by 
the SNR age of 1,600 years \citep{rob03}.
The temperature of the  X-ray emitting gas is 0.58 keV, 
so that its pressure becomes 
$p/k_B$ = 1.8 $\times$ 10$^8$ cm$^{-3}$ K,
where $k_B$ is the Boltzmann constant.
For comparison,
the average electron densities of clumps are 4,700--9,400 cm$^{-3}$. Using the
characteristics temperature of 5,000 K, 
we get the ionized gas pressure 
$(n_e+n_p)~T$ = 0.47--0.94 $\times$ 10$^8$ cm$^{-3}$ K,
which is about 1/4 to 1/2 of the pressure from the X-rays,
indicating that the ionization fraction of \feii\ emitting gas is low
\citep[e.g.,][]{oli89}.
The total gas pressure, neglecting He, is 
$p/k_B=(1+f_e)/(2f_e) \times n_e T_e$, 
where $f_e$, is the mean ionization fraction. 
This gives $f_e=0.14$--0.34, 
consistent with our previous results from the shock model calculations 

The southeastern shell of G11.2$-$0.3 has been detected 
by Spitzer mid-infrared IRAC 3--8 $\mu$m and MIPS 24 $\mu$m images \citep{lee05, rea06, pin11}.
Comparing the distributions of the mid-infrared images with those of the near-infrared
\feii\ and H$_2$ line images \citep{koo07},
the brightest Clump~1 is less pronounced in the mid-infrared images,
and a small molecular cloud located at $\sim$ 1\arcmin\ southwest from Clump~1 
(H$_2$-pk1 in \citealt{koo07})
is prominent in the Spitzer IRAC images.
This is consistent with the classification of G11.2$-$0.3 (H$_2$-pk1)
with molecular shocks
by the ratios of the IRAC intensities \citep{rea06},
and the Spitzer mid-infrared spectrum,
obtained at the position of this molecular cloud,
shows a wealth of H$_2$ pure rotational and ionic lines 
as well as bright dust continuum \citep{and11}.
Unfortunately, the position of the molecular cloud is
outside the \feii\ clumps and not covered by our observations.
For the Spitzer MIPS 24 $\mu$m emission of Clump~1, although we 
expect a significant contribution from \feii\ 24.52 and 25.99 $\mu$m lines,
it is not straightforward to determine the relative contributions from these 
ionic lines and dust continuum, due to various uncertainties in comparing
the model spectrum to the observed brightness.
Future imaging spectroscopic observations are needed for detailed study of the mid-infrared emission.

An interesting finding of our observations is the detection 
of SE-BHVF which is Doppler shifted by $-$440 $\pm$ 100 km s$^{-1}$ 
around the northeastern corner of Clump~1 (Figures~\ref{fig_limg} and \ref{fig_sp}c).
Assuming that SE-BHVF is located at the boundary of 2\arcmin\ radius of the SNR shell 
with its projected distance of 1.7\arcmin\ from the center,
its de-projected expanding velocity is $\sim$~$-$900~km~s$^{-1}$ 
with respect to the +45 \kms\ systematic velocity of \gele\ \citep{gre88}.
This expanding velocity is comparable to the velocities 
of the highly blueshifted ($\sim$ $-$1000 km s$^{-1}$) ejecta knots 
previously detected around the center of \gele\ \citep{moo09},
which is suggestive that SE-BHVF has the same origin as the central knots.
If \gele\ is indeed the remnant of the supernova explosion in AD 386 \citep{cla77},
the observed ejecta velocities indicate that they have been considerably decelerated.
The existence of SE-BHVF in Clump 1 of the southeastern shell
is consistent with the previous interpretations that the SN ejecta in \gele\
was expelled mainly in the southeast-northwest direction, 
given the existence of the northwestern filaments 
which are most likely of the ejecta origin \citep{koo07, moo09}.

\section{Conclusion}

We perform near-infrared (1.0--1.8 $\mu$m) integral-field spectroscopic observations of 
the southeastern shell of the young core-collapse SNR G11.2$-$0.3
using our own instrument setup that provides an unprecedented integral-field size in the wavebands.
The shell is known to have brightest near-infrared \feii\ emission
of all Galactic SNRs, and is presumed to be a unique mixture of 
progenitor's circumstellar material and supernova ejecta. 
Our data cube populated with multiple \feii\ transitions 
makes it possible to study the detailed two-dimensional distributions of 
the line intensities, velocities, electron densities, and extinction.
Based on this information, we compare the observed results with those
predicted by SNR shock models and obtain shock parameters in this region of the SNR.
The observed properties of the brightest clump can be explained 
by multiple shocked thin filaments of circumstellar material.
This, together with the discovery of the faint high-velocity component,
suggests that the southeastern shell of \gele\ is indeed a mixture
of circumstellar material and ejecta, 
with the former being the dominant component.
Future studies with improved sensitivity and resolutions 
(both spatial and spectral) are needed to understand how the two components are interacting.

This work is supported by a Grant-in-Aid for Japan Society of Promotion of Science (JSPS) fellows
(No. 23$\cdot$01322). 
H.-G.L. was partly supported from Early Research Award to D.-S.M. by the Ministry of Economic
Development and Innovation (MEDI) of the Ontario Provincial Government in Canada.
D.-S.M. acknowledges support from the Natural Science and Engineering Research Council (NSERC) of Canada. 
This paper was studied with the support of the Ministry of Education Science and Technology (MEST) 
and the Korean Federation of Science and Technology Societies (KOFST).
B.-C.K. was supported by the National Research Foundation of
Korea (NRF) Grant (NRF-2010-616-C00020).
FISICA was supported by the Florida Space Research Initiative.

{\it Facility:} \facility{Mayall (FLAMINGOS)}

{}

\clearpage

\clearpage
\begin{deluxetable}{lc cc cc cc}
\tabletypesize{\scriptsize}
\tablewidth{0pt}
\tablecolumns{8} 
\tablecaption{Measured line intensities of Clump 1, Clump 2, and Clump 3
\label{tab_obs}}
\tablehead {
\multicolumn{2}{c}{Line}  &\multicolumn{2}{c}{Clump 1} &\multicolumn{2}{c}{Clump 2} &\multicolumn{2}{c}{Clump 3} \\
\colhead{Transition} &\colhead{$\lambda$ ($\mu$m)}
&\colhead{Observed} &\colhead{Dereddened} 
&\colhead{Observed} &\colhead{Dereddened}
&\colhead{Observed} &\colhead{Dereddened}
}
\startdata
~                 He I $^3S_1 \rightarrow ^3P$     &   1.083     &   3.5 $\pm$   0.8     &  50.0 $\pm$  11.2     &                       &                       &                       &                      \\
~[Fe II] $a^6 D_{9/2} \rightarrow a^4 D_{7/2}$     &   1.257     &  30.3 $\pm$   1.9     & 136   $\pm$   8.6     &  25.9 $\pm$   2.1     & 136   $\pm$  11.2     &  22.5 $\pm$   1.9     & 136   $\pm$  11.4    \\
~[Fe II] $a^6 D_{5/2} \rightarrow a^4 D_{5/2}$     &   1.295     &   3.4 $\pm$   0.8     &  12.7 $\pm$   3.1     &   3.5 $\pm$   1.2     &  14.7 $\pm$   5.2     &                       &                      \\
~[Fe II] $a^6 D_{7/2} \rightarrow a^4 D_{7/2}$     &   1.321     &  15.2 $\pm$   1.4     &  49.4 $\pm$   4.4     &  11.2 $\pm$   2.0     &  41.0 $\pm$   7.2     &   9.8 $\pm$   2.0     &  40.0 $\pm$   8.4    \\
~[Fe II] $a^6 D_{3/2} \rightarrow a^4 D_{5/2}$     &   1.328     &   5.7 $\pm$   1.4     &  17.8 $\pm$   4.3     &                       &                       &                       &                      \\
~[Fe II] $a^4 F_{9/2} \rightarrow a^4 D_{5/2}$     &   1.534     &  13.7 $\pm$   0.9     &  19.2 $\pm$   1.2     &  12.8 $\pm$   2.4     &  18.5 $\pm$   3.4     &   8.5 $\pm$   0.9     &  12.7 $\pm$   1.4    \\
~[Fe II] $a^4 F_{7/2} \rightarrow a^4 D_{3/2}$     &   1.600     &   6.1 $\pm$   0.9     &   7.0 $\pm$   1.0     &   9.0 $\pm$   1.7     &  10.4 $\pm$   2.0     &   8.5 $\pm$   1.9     &   9.9 $\pm$   2.2    \\
~[Fe II] $a^4 F_{9/2} \rightarrow a^4 D_{7/2}$     &   1.644     & 100   $\pm$   5.1     & 100   $\pm$   5.1     & 100   $\pm$   9.7     & 100   $\pm$   9.7     & 100   $\pm$   5.5     & 100   $\pm$   5.5    \\
~[Fe II] $a^4 F_{5/2} \rightarrow a^4 D_{1/2}$     &   1.664     &   4.4 $\pm$   1.1     &   4.1 $\pm$   1.0     &   4.7 $\pm$   1.3     &   4.4 $\pm$   1.2     &                       &                      \\
~[Fe II] $a^4 F_{7/2} \rightarrow a^4 D_{5/2}$     &   1.677     &   8.9 $\pm$   0.6     &   8.2 $\pm$   0.6     &  11.6 $\pm$   1.8     &  10.5 $\pm$   1.6     &   9.3 $\pm$   0.9     &   8.3 $\pm$   0.8    \\
~[Fe II] $a^4 F_{5/2} \rightarrow a^4 D_{3/2}$     &   1.712     &   3.4 $\pm$   0.4     &   2.8 $\pm$   0.4     &                       &                       &                       &                      \\
~[Fe II] $a^4 F_{3/2} \rightarrow a^4 D_{1/2}$     &   1.745     &   3.8 $\pm$   0.6     &   2.9 $\pm$   0.5     &                       &                       &                       &                      \\
~[Fe II] $a^4 F_{3/2} \rightarrow a^4 D_{3/2}$     &   1.798     &   6.0 $\pm$   1.2     &   4.1 $\pm$   0.8     &   9.7 $\pm$   2.9     &   6.4 $\pm$   1.9     &                       &                      \\
\enddata
\tablecomments{
Normalized to make the peak intensity of [Fe II] 1.644 \um\ line of each clump 100.
}
\end{deluxetable}

\end{document}